\documentclass[12pt]{article}
\usepackage{amsmath,amssymb,amsthm,amscd}
\input epsf.sty
\newcommand{\CK}{{\cal K}}
\newcommand{\CL}{{\cal L}}

\newcommand{\CN}{{\cal N}}
\newcommand{\CO}{{\cal O}}

\def\IZ{{\mathbb Z}}

\def\IC{{\mathbb C}}
\def\IP{{\mathbb P}}

\def\IS{{\mathbb S}}

\newcommand{\tr}{{\rm Tr}}
\newcommand{\re}{{\rm e}}
\newcommand{\ri}{{\rm i}}
\newcommand{\rd}{{\rm d}}

\newcommand{\be}{\begin{equation}}
\newcommand{\ee}{\end{equation}}
\newcommand{\ba}{\begin{aligned}}
\newcommand{\ea}{\end{aligned}}
\newcommand{\ben}{\begin{eqnarray}\displaystyle}
\newcommand{\een}{\end{eqnarray}}

\newcommand{\sectiono}[1]{\section{#1}\setcounter{equation}{0}}

\newdimen\tableauside\tableauside=1.0ex
\newdimen\tableaurule\tableaurule=0.4pt
\newdimen\tableaustep
\def\phantomhrule#1{\hbox{\vbox to0pt{\hrule height\tableaurule width#1\vss}}}
\def\phantomvrule#1{\vbox{\hbox to0pt{\vrule width\tableaurule height#1\hss}}}
\def\sqr{\vbox{%
  \phantomhrule\tableaustep
  \hbox{\phantomvrule\tableaustep\kern\tableaustep\phantomvrule\tableaustep}%
  \hbox{\vbox{\phantomhrule\tableauside}\kern-\tableaurule}}}
\def\squares#1{\hbox{\count0=#1\noindent\loop\sqr
  \advance\count0 by-1 \ifnum\count0>0\repeat}}
\def\tableau#1{\vcenter{\offinterlineskip
  \tableaustep=\tableauside\advance\tableaustep by-\tableaurule
  \kern\normallineskip\hbox
    {\kern\normallineskip\vbox
      {\gettableau#1 0 }%
     \kern\normallineskip\kern\tableaurule}%
  \kern\normallineskip\kern\tableaurule}}
\def\gettableau#1{\ifnum#1=0\let\next=\null\else
\squares{#1}\let\next=\gettableau\fi\next}

\newcommand{\figref}[1]{Fig.~\protect\ref{#1}}
\begin{document}
\begin{titlepage}
{}~
\hfill\vbox{
%\hbox{hep-th/0612127}
\hbox{CERN-PH-TH/2006-258}
}\break

\vskip .6cm

\centerline{\Large \bf
Open string amplitudes and large order behavior}
\vspace*{1.0ex}
\centerline{\Large \bf in topological string theory}

\medskip

\vspace*{4.0ex}

\centerline{\large \rm
Marcos Mari\~no\footnote{Also at Departamento de Matem\'atica, IST, Lisboa, Portugal}}

\vspace*{4.0ex}

\centerline{  Department of Physics, CERN}
\centerline{  Geneva 23, CH-1211 Switzerland}
\vspace*{2.0ex}
\centerline{marcos@mail.cern.ch}

\vspace*{15.0ex}

\centerline{\bf Abstract} \bigskip

We propose a formalism inspired by matrix models to compute open and closed 
topological string amplitudes in the B-model on toric Calabi--Yau manifolds. 
We find closed expressions for various open string amplitudes beyond the disk, 
and in particular we write down the annulus amplitude in terms of  theta functions on a Riemann surface. 
We test these ideas on local curves and local surfaces, providing 
in this way generating functionals for open Gromov--Witten invariants in the spirit of mirror symmetry. In the 
case of local curves, we study the open string sector near the critical point which leads to 2d gravity, and we show that 
toric D-branes become FZZT branes in a double-scaling limit. We use this connection to compute non-perturbative 
instanton effects due to D-branes that control the large order behavior of topological string theory on these backgrounds.

\end{titlepage}

\baselineskip=16pt

\tableofcontents

\sectiono{Introduction}

Topological strings constitute an important subsector of string theory with various physical and mathematical 
applications, and they have been extensively investigated since they were first formulated. This has led to many different ways 
of computing their amplitudes based very often on string dualities. Topological strings come in two types, the A model and 
the B model, which are related by mirror symmetry. The A model provides a physical formulation of Gromov--Witten theory, and it can 
be solved on toric backgrounds by a large $N$ duality with Chern--Simons theory. The B model is deeply 
related to the theory of deformation of complex structures, and the amplitudes in the closed string sector 
satisfy holomorphic anomaly equations \cite{bcov} which can be used to solve the theory in many cases. 

The open string sector of topological string theory is not as well understood as the closed one. 
For example, the open string analogue of the holomorphic anomaly equations 
has not been formulated yet. Most of the explicit results in the open B model refer to disk amplitudes, which have been 
studied in detail in the local Calabi--Yau (CY) case \cite{av,akv,india,lm,lmw} and more recently in the compact case \cite{walcher,psw}. In \cite{adkmv} a general 
framework was proposed to understand the B--model on local CY geometries, but in practice its use has been limited to the simplest local model i.e. 
the topological vertex.  

On top of these practical problems concerning the computation of the amplitudes, there are more conceptual problems related to the 
nature of the string perturbation series. Like other string theories, topological strings are defined in a perturbative, genus by genus expansion, 
and we do not have a general definition beyond perturbation theory. Some 
proposals have emerged based on holographic descriptions related to black hole physics \cite{osv}, and in some simple cases 
these descriptions have been worked out in detail \cite{aosv}, but much remains to be done in order to understand these fundamental 
issues. 

In this paper we discuss the open string sector of the B model and the nature of the topological string perturbation series by relying on ideas and techniques of 
matrix models. Dijkgraaf and Vafa have shown that, in some local Calabi--Yau backgrounds, the B model reduces to a matrix model \cite{dv}, but the backgrounds 
they considered do not include the toric backgrounds with an interesting enumerative geometry. Some of these, like the resolved conifold and the 
$A_N$ fibrations over $\IP^1$ that engineer $\CN=2$ gauge theories, admit a matrix model description closely related to Chern--Simons theory 
\cite{m,akmv}. However, the matrix model that one finds in this way
is difficult to solve and its applications have been limited.  

A way to remedy this situation is to notice that the $1/N$ expansion of a matrix model can be substantially generalized and 
reformulated, leading to a geometric construction that starts with a Riemann surface 
and defines recursively a set of open and closed amplitudes on it. This 
construction emerges in a natural way when one solves the loop 
equations of a matrix model. It was started in \cite{ackm} and has culminated in the work of Eynard and collaborators 
(see \cite{eynard, ce, eco}, and specially \cite{eo}, where this point of view is stated most clearly). The general construction 
includes as a particular case the $1/N$ expansion of the standard one--matrix model with a polynomial potential. In that case, the 
spectral curve encoding the planar solution can be described by an equation of the form, 
\be
\label{ycurve}
y^2(x)=M^2(x) \sigma (x), \quad \sigma(x) =\prod_{i=1}^n (x-x_i),
\ee
where $M(x)$ is a polynomial in $x$. The corresponding closed and open amplitudes obtained from the general geometric construction of 
\cite{eynard, eo} are then the free energies and correlation functions, respectively, of the matrix model underlying (\ref{ycurve}). 

For an arbitrary Riemann surface 
we should expect that the quantities obtained in this formalism are general amplitudes for a chiral boson living on the 
curve, since the $1/N$ expansion of a matrix model can be reformulated in this way \cite{kostov}. On the other hand, 
it was shown in \cite{adkmv} that the B model on a 
general local Calabi--Yau geometry is described as well by a chiral boson living on a Riemann surface. 
It is then natural to conjecture that the geometric amplitudes defined in \cite{eynard,eco, eo} 
describe open and closed string amplitudes for the local B model, once the appropriate spectral curve
has been found. We will see in this paper that this is the case also for toric geometries. The spectral curve can be obtained 
from the disk amplitudes, and it is of the form (\ref{ycurve}) but with a nonpolynomial $M(x)$. Using this we 
give closed B model formulae for various open string amplitudes on toric CY threefolds beyond the disk level, like annulus amplitudes and three--holed sphere 
amplitudes, which we test in various examples. In these tests 
we focus on so--called outer brane amplitudes of toric geometries. We expect that our formalism can accommodate more general cases, 
and we offer a preliminary discussion of this issue in the concluding section. 

Another advantage of matrix models is that one has a good control of instanton effects in the form of eigenvalue tunneling \cite{david, shenker}. This has been used to 
understand the properties of the string perturbation series in the double--scaled models which describe non--critical strings, by using the standard connection 
between instanton effects and large order behavior in perturbation theory \cite{zj}. In \cite{it} it was found that topological strings on local curves provide a realization of 2d gravity near a critical point 
at small radius in K\"ahler moduli space. This was done through a detailed analysis of the closed string sector. Here we analyze the open string sector of these models 
to test our general matrix model formalism, and we show that, in the double scaling limit, toric D--branes become the FZZT branes of 2d gravity. 
This allows us to identify computable, non-perturbative instanton effects due to D-branes which are responsible for the large order behavior of topological 
string perturbation theory away from criticality. Our analysis suggests that in some circumstances one can find Borel--summable string perturbation series. 

The organization of this paper is the following. In section 2 we review open string amplitudes in the local setting and the relevant matrix model 
results, and we present our proposal for the computation of open (and closed) string amplitudes in the local B model. In section 3 we analyze the example 
of local curves. We test our proposal and we analyze the instanton effects controlling the large order behavior. Section 4 presents further tests for 
two local surfaces: local $\IP^1 \times \IP^1$ and local $\IP^2$. In section 5 we conclude with various problems and comment on possible developments. 

\sectiono{Open string amplitudes and matrix models}

In this section we first review various aspects of topological string theory relevant to this paper and then 
we make our proposal. General recent references on topological 
strings include \cite{mm,vonk,nv}, to which we refer the reader for background material. 

\subsection{General aspects and open Gromov--Witten interpretation}

Let us consider topological string theory on a Calabi--Yau manifold $X$. To define the open string sector in the A model we have 
to speciy a Lagrangian submanifold $\CL$ of  $X$ \cite{witten}. The type-A open 
topological string theory describes holomorphic maps from open Riemann 
surfaces of genus $g$ and with $h$ holes $\Sigma_{g,h}$ to the Calabi--Yau $X$, with Dirichlet boundary 
conditions specified by $\CL$. These holomorphic maps are usually called open string instantons. The
topological sector of an open string instanton is given by two different kinds of data:
the boundary part and the bulk part. For the bulk part, the topological
sector is labelled by relative homology classes, since we are requiring the
boundaries of $f_*[\Sigma_{g,h}]$ to end on $\CL$. Therefore, we will
set
\begin{equation}
f_*[\Sigma_{g,h}]=\beta \in H_2(X, \CL).
\label{bulkpart}
\end{equation}
To specify the topological sector of the boundary, we will assume that
$b_1 (\CL)=1$, so that $H_1 (\CL)$ is generated by a non-trivial
one-cycle $\gamma$. We then have
\begin{equation}
f_*[C_i]=w_i \gamma, \,\,\,\,\, w_i \in {\bf Z},\,\,\,\,
i=1, \cdots, h,
\label{wind}
\end{equation}
i.e. $w_i$ is the winding number associated to the map $f$
restricted to $C_i$. We will collect these integers into a single
$h$-uple denoted by $w=(w_1, \cdots, w_h)$.

The free energy of type-A open topological string theory at fixed genus and 
boundary data $w$, which we denote by $F_{w,g}(t)$, can be computed as a sum over open string instantons 
labelled by the bulk classes: 
\begin{equation}
F_{g,w} (t) =\sum_{\beta} F_{g,w,\beta}\,Q^{\beta}.
\label{openGW}
\end{equation}
In this equation, the sum is over relative homology classes $\beta \in H_2(X,\CL)$. The quantities $F_{w,g, \beta}$
are open Gromov--Witten invariants, and they `count' in an appropriate sense the number of holomorphically
embedded Riemann surfaces of genus $g$ in $X$ with Lagrangian boundary
conditions specified by $\CL$, and in the class represented
by $\beta, w$. Finally, $Q^{\beta}$ denotes $\exp(-\sum_i \beta_i t_i)$, where the index $i=1, \cdots, b_2(X,\CL)$ runs over a basis of 
$H_2(X,\CL)$. It is useful to have a generating functional for open string amplitudes at fixed genus
\be
\label{fgz}
F_g (z_1, \cdots, z_h)=\sum_{w_i} F_{g,w} (t) z_1^{w_1} \cdots z_h^{w_h}.
\ee
Particularly important cases of these amplitudes are the disk and the annulus amplitudes, which correspond to $g=0$ and $h=1,2$, respectively. We will denote 
them by $W(z)$ and $A(z_1, z_2)$. These amplitudes have the following spacetime interpretation \cite{bcov,mayr}. If one considers $N$ D6 branes wrapping the Lagrangian 
submanifold $\CL$, the resulting four--dimensional worldbrane theory is an $U(N)$ gauge theory with $\CN=1$ supersymmetry and chiral fields associated 
to the closed and open moduli $t$, $z$. The superpotential for these moduli is precisely the disk instanton amplitude $W(z)$, while $A(z,z)$ gives the gauge kinetic term. 

It should be possible to compute these open string amplitudes directly in the B-model, and then use mirror symmetry to relate 
them to the amplitudes in the A model, as it was done in \cite{av,akv} for the disk. A special case where we know the B model answer is 
the situation considered by Dijkgraaf and Vafa in \cite{dv}. In the local CY backgrounds of \cite{dv} the B-model reduces 
to a matrix model for a Hermitian matrix $M$, and open string amplitudes have the following interpretation. Define the connected $h$--point functions, 
\be
\label{mmcorr}
W(p_1,\cdots, p_h)=\big\langle \tr\, {1\over p_1-M} \cdots \tr {1\over p_h-M}\big\rangle^{(c)},
\ee
which are generating functionals for matrix model correlators
\be
W(p_1, \cdots, p_h) =\sum_{k_i\ge 1} \langle \tr \, M^{k_1} \cdots \tr \, M^{k_h} \rangle^{(c)} p_1^{-k_1-1} \cdots p_h^{-k_h-1}.
\ee
These quantities have a genus expansion of the form,
\be
W(p_1, \cdots, p_h) =\sum_{g=0}^{\infty} g_s^{2g-2+h} W_g(p_1, \cdots, p_h).
\ee
If we set $z_i=p_i^{-1}$, the relation with (\ref{fgz}) is simply
\be
W_g(z_1, \cdots, z_h) =z_1^2 \partial_{z_1} \cdots z_h^2 \partial_{z_h}  F_g(z_1, \cdots, z_h), \quad h\ge 2.
\ee
The case of $g=0, h=1$ is slightly different. The generating functional is the resolvent of the 
matrix model, and it is related to the disk amplitude as 
\be
\label{resdisk}
\omega_0(z) =z + z^2 \partial_z W(z).
\ee
In the case of a one--matrix model with a polynomial potential $V(M)$, the resolvent  is always of the form (see for example \cite{dfgz,mmhouches})
\be
\label{omsigma}
\omega_0(p)={1\over 2} (V'(p) + y (p)), 
\ee
where $y(p)$ is the spectral curve describing the planar model and can be written as
in (\ref{ycurve}). The points $x_i$ are the branch points of the curve.  

Consider now the case in which $X$ is a toric CY threefold. The mirror CY can be written in terms of an algebraic equation,
\be
\label{curve}
F(\re^u,\re^v)=0,
\ee
which defines implicitly a function $v(u)$. It was shown in \cite{av} that in these geometries there are 
Lagrangian submanifolds $\CL$ with the topology of $\IS^1 \times \IC$ whose disk amplitudes can be easily computed in the mirror 
geometry (\ref{curve}). As shown in \cite{av, akv}, they are simply given by 
\be
\label{disk}
W(u)=\int^u \rd u'\, v(u').
\ee
Here, one identifies the open string modulus as $z=\re^u$. There are various subtleties in this procedure. First of all, there are various 
types of Lagrangian submanifolds in these geometries depending on their location in the toric diagram. The choice of parametrization 
of the curve (\ref{curve}) depends on this location. Second, the variables appearing in (\ref{curve}) are not flat coordinates. There is an open/closed 
mirror map of the form
\be
U =U(u,t), 
\ee
where the flat open coordinate depends on the closed string moduli $t$. 
One has to reexpress the superpotential in terms of the $U$ variable in order to extract geometric information.  We will 
discuss these issues in more detail in section 4, when we consider concrete examples. In any case, the above results suggest to define a resolvent for these local models as
\be
\label{resolvent}
\omega_0(p)= {1\over p}(1 +v(p)),
\ee
The function $v(p)$ has in general the structure
\be
\label{vfunction}
v(p)=\log \biggl[ {a(p) + {\sqrt {\sigma(p)}} \over c(p)}\biggr], \quad \sigma(p) =\prod_{i=1}^n (p-x_i).
\ee
If we now use the identity, 
\be
\label{logtanhidentity}
\log \biggl( {a + {\sqrt {\sigma}} \over c}\biggr)= {1\over 2} \log {a^2-\sigma \over c^2} + \tanh^{-1}\Bigl({{\sqrt {\sigma}}  \over a}\Bigr).
\ee
we see, by comparing (\ref{omsigma}) and (\ref{vfunction}), that the role of the spectral curve is played in these geometries by the equation
\be
\label{toricsc}
y(p)={2 \over p} \tanh^{-1}\biggl[ {\sqrt{ \prod_{i=1}^n (p-x_i)} \over a(p)}\biggr].
\ee
This can be written in the form (\ref{ycurve}) but with a non--polynomial function $M(p)$:
\be
M(p)={2\over p {\sqrt {\sigma (p)}}}  \tanh^{-1}\biggl[ {\sqrt{ \sigma (p)} \over a(p)}\biggr].
\ee
Notice that this curve can be obtained from the generating function for disk amplitudes $v(p)$, or equivalently from the D6 brane superpotential $W(z)$. 

Finally, we recall that open topological string amplitudes on Calabi--Yau manifolds have an underlying integrality structure first obtained in \cite{ov} and 
further refined in \cite{lmv,mv}. In the case of genus zero amplitudes with $h$ holes, integrality leads to an expansion of the form 
\be
\label{integers}
F_{0,w}(z_1, \cdots, z_h)= \sum_{\beta, w_i}\sum_{d=1}^{\infty} d^{h-3} n_{0,w}^{\beta} z_1^{dw_1} \cdots z_h^{d w_h}Q^{d\beta},
\ee
where $n_{0,w}^{\beta}$ are integer numbers and give open string analogues of the Gopakumar--Vafa invariants.

\subsection{The annulus amplitude}

We start our discussion of open string amplitudes with the annulus amplitude, 
\be
A(p,q)=\sum_{k,\ell\ge 1} F_{0,k,\ell} (t)p^{-k} q^{-\ell},
\ee
where $k,\ell$ are winding numbers. It corresponds to the two--point planar amplitude 
in the matrix model,
\be
\label{bk}
W_0(p,q)=\partial_p \partial_q A(p,q),
\ee
with the expansion, 
\be
\label{wexp}
W_0(p,q) = \sum_{k,\ell \ge 1} k \ell F_{0,k,\ell}(t) p^{-k-1} q^{-\ell -1}.
\ee
The amplitude $W_0(p,q)$ is, together with the spectral curve itself, the most 
important ingredient in the analysis of loop equations, since it determines the rest of the amplitudes by recursion. 
In the case of the annulus amplitude it is also easy to make the connection between the theory of the chiral boson and 
the formalism of matrix models \cite{kostov}. 

Let us then compute the annulus amplitude in the B--model, in the case of a local Calabi--Yau case described by (\ref{curve}). As shown in \cite{adkmv}, 
the B--model on (\ref{curve}) reduces to the theory of a chiral boson living on the curve $\Sigma_g$ associated to (\ref{curve}). This chiral boson has a classical 
piece given by 
\be
\partial \phi_{\rm cl}(u)=v(u).
\ee
The disk amplitude is given by the one--point function 
of this chiral boson, 
\be
W(u) =\langle \phi (u)\rangle.
\ee
Since only the classical part of the boson field contributes to this correlation function, one immediately recovers (\ref{disk}). 
The annulus amplitude should be given by the connected two--point function of the chiral boson, 
\be
\label{anncorr}
A(p,q)=\langle \phi (p) \phi(q) \rangle^{(c)}.
\ee
Here, $p,q$ are local coordinates on the Riemann surface chosen in such a way that, as $p,q \rightarrow \infty$, $A(p,q)$ vanishes. 
Only the quantum fluctuations of the chiral boson contribute to this connected correlator. It is well--known that 
this correlation function is a natural object on the Riemann surface $\Sigma_g$ which can be expressed in terms of 
theta functions. Let us define the Abel map,
\be
u_i(p)=\int_{p_0}^p  \omega_i (p') \rd p', \quad i=1, \cdots, g,
\ee
where $\omega_i (p) \rd p$ is a basis of Abelian differentials on $\Sigma_g$ and $p_0$ is a 
basepoint. We then have, 
\be
\label{anntheta}
A(p,q)=\log \, \vartheta_* ({\bf u}(p) -{\bf u}(q)|\tau) -\log \, (p-q). 
\ee
where the subscript $*$ means that the theta function has an {\it odd} characteristic, and we have subtracted the 
singular part of the correlator. The closely related quantity (\ref{bk}) is given by
\be
\label{twopoint}
W_0(p,q)=\langle \partial \phi (p) \partial \phi(q) \rangle^{(c)}= B(p,q) -{1\over (p-q)^2}
\ee
where $B(p, q)$ is sometimes called the Bergmann kernel. It can be also written as 
\be
\label{bergmann}
B(p,q)=\partial_p \partial_q \log \, E(p,q),
\ee
where $E(p,q)$ is the prime form on $\Sigma_g$. Equation (\ref{anntheta}) generalizes the computation of the annulus amplitude for a target torus in \cite{mirror}, chapter 35.4, to mirror geometries given by a general 
Riemann surface $\Sigma_g$. The result (\ref{twopoint}) for the two--point connected function is a standard result in the matrix model literature, see for example \cite{eynard,eo}.

As it stands, the formula (\ref{anntheta}) is not easy to use in explicit computations. Fortunately, the Bergmann kernel $B(p,q)$ 
plays an important role in the theory of matrix models since it gives an explicit expression for the correlation function (\ref{anncorr}) 
(see for example \cite{eynard,eo} and references therein). Its most 
important property is that it only depends on the spectral curve through the positions of the branch points $x_i$. We will assume in the following 
that there are $2s$ branch points coming in pairs. The fact that $W_0(p,q)$ only depends on $x_i$ was proved in \cite{ajm} for $s=1$, in \cite{akemann} 
for $s=2$, and for general $s$ in \cite{aa}. For $s=1$ the spectral curve has genus zero and $W_0(p,q)$ is given by \cite{ajm}
\be
\label{onecutannulus}
W_0(p,q)= {1\over 2 (p-q)^2} \biggl( {p q -{1\over 2}(p+q) (x_1+x_2) +x_1 x_2 \over {\sqrt{(p-x_1)(p-x_2)(q-x_1)(q-x_2)}}}-1\biggr).
\ee
For $s=2$ the spectral curve has genus one, and $W_0(p,q)$ can be written in terms of Weierstrass function (see, for example, section 3.6 of 
\cite{bde}). An explicit expression in terms of the branch points was found by Akemann in \cite{akemann}:
\be
\label{twocutannulus}
\ba
W_0(p,q)&={1\over 4(p - q)^2} \biggl(   {\sqrt  {(p - x_1) (p - x_4)(q -x_2) (q - x_3) \over (p - x_2) (p - x_3)(q - x_1) (q - 
                      x_4)}} \\
                      & \, \, \ \ \ \ \ \ \ \ \ \ \ \ +   {\sqrt  {(p - x_2) (p - x_3)(q -x_1) (q - x_4) \over (p - x_1) (p - x_4)(q - x_2) (q - 
                      x_3)}} \biggr) \\ & 
                      + {(x_1-x_3)(x_2-x_4) \over 4 {\sqrt{\prod_{i=1}^4 (p-x_i)(q-x_i)}}}  {E(k) \over K(k)}  - {1\over 2
                      (p - q)^2},
                      \ea
\ee
where $K(k)$, $E(k)$ are elliptic functions of the first and second kind with modulus
\be
\label{kpar}
k^2={(x_1-x_4)(x_2-x_3) \over (x_1-x_3)(x_2-x_4)}.
\ee
It can be easily checked that these expressions have the expansion in $p, q$ given in (\ref{wexp}). One finds, for example, that
\be
F_{0,1,1}(t)={(x_1-x_2)^2 \over 16}, \quad F_{0,1,2}(t)={1\over 32}(x_1-x_2)^2(x_1+x_2), 
\ee
for $s=1$. For $s=2$ one has 
\be
F_{0,1,1}(t)={1\over 4}(x_1-x_3)(x_2-x_4){E(k) \over K(k)}  +{1\over 16} (x_1 + x_4-x_2 -x_3)^2.
\ee
In order to apply these formulae in the toric case, we write the spectral curve (\ref{toricsc}) from the disk amplitudes, and we read the position of the 
branch points from the zeros of the square root involved in the expression. How do we order them? The branch points come in pairs and define cuts in the 
complex plane. Each cut corresponds to a fixed  ``filling fraction." For example, for $s=2$ the cuts are 
$(x_1, x_4)$ and $(x_2, x_3)$ \cite{kmt}. In the B model context, we will require that the cuts shrink to zero size at large radius/large complex 
structure, so that $k^2=0$ at this point. This will lead to an expansion of the amplitudes compatible with the A-model interpretation at large radius.

\subsection{Holomorphic anomaly and Ray--Singer torsion}

The annulus amplitudes $A(p,q)$, $W_0(p,q)$ defined by (\ref{anntheta}), (\ref{twopoint}) are holomorphic, but they are
not modular invariant. Under a symplectic transformation
\be
\tau \rightarrow (A \tau + B)(C\tau + D)^{-1}, 
\ee
the theta function transforms as
\be
\vartheta_* ({\bf u}|\tau) \rightarrow \Bigl( {\rm det}(C \tau +D)\Bigr) ^{1\over 2} \exp[\ri \pi {\bf u} (C\tau + D)^{-1}C {\bf u}] \vartheta_*({\bf u}|\tau),
\ee
and $W_0(p,q)$ transforms with a shift\footnote{The modular properties of open and closed amplitudes in matrix models are extensively discussed in \cite{eo}.} 
\be
W_0(p,q) \rightarrow W_0(p,q) - 2 \pi \ri  {\bf \omega}(p)(C\tau + D)^{-1}C {\bf \omega}(q).
\ee
Following the philosophy of \cite{cardoso,hk, abk}, one can obtain a modular invariant object by adding an antiholomorphic piece in the 
closed string moduli, 
\be
\label{tw}
\widetilde W_0(p,q) =W_0(p,q) - \pi  {\bf \omega}(p) ({\rm Im}\, \tau)^{-1}  {\bf \omega}(q).
\ee
Notice that this object is still holomorphic as a function of the open string moduli $p$, $q$, and satisfies a very simple holomorphic anomaly equation 
for the closed string moduli. The K\"ahler metric on the Jacobian is 
\be
g_{k \bar l}=-\ri (\tau -\bar \tau)_{k l}, 
\ee
and the tensor
\be
C_{ijk}={\partial \tau_{ij}\over \partial t^k}
\ee
is the three--point function in the closed string sector. The antiholomorphic dependence of $\widetilde W_0(p,q)$ is captured by 
\be
\label{holan}
{\partial \widetilde W_0(p,q) \over \partial \bar t^{\bar k}} =2\pi \ri \omega_i(p) {\overline C}_{\bar k}^{ij} \omega_j(q), 
\ee
where we have raised the indices with the inverse metric $g^{i \bar j}$. In the context of the holomorphic anomaly equation of \cite{bcov} one 
can construct a propagator $S^{ij}$ satisfying 
\be
\partial_{\bar k} S^{ij} = {\overline C}_{\bar k}^{ij},
\ee
therefore one can integrate (\ref{holan}) to write
\be
\widetilde W_0(p,q)=2\pi \ri \omega_i(p) S^{ij} \omega_j(q) + f, 
\ee
where $f$ (the ambiguity appearing as an integration constant) is holomorphic in the closed string moduli. 

It was shown in \cite{bcov} that in the B-model the annulus amplitude should be related to the Ray--Singer torsion of the $\bar \partial$ operator 
coupled to a vector bundle $V$ \cite{rs}. This bundle is associated to the D-branes that provide the boundary conditions \cite{witten}. In fact, the Ray--Singer 
torsion should provide a natural modular invariant completion of the 
annulus amplitude (\ref{anntheta}). We now discuss the relation between both quantities. The relevant background on differential operators on Riemann surfaces can 
be found in \cite{agmv,dhp}.

Flat line bundles over a Riemann surface $\Sigma_g$ are parametrized by the Jacobian $J(\Sigma_g)$. Let us then consider the line bundle 
$\CL$ associated to 
\be
{\bf v}={\bf u}(p) -{\bf u}(q) \in {\rm Jac}(\Sigma_g). 
\ee
In terms of divisors, this is the line bundle 
\be
\CL=[p-q].
\ee
We now consider the Cauchy--Riemann operator 
\be
\bar\partial_{\CL}:\Omega(\CL) \rightarrow \Omega^{0,1}(\CL).
\ee
The Ray--Singer torsion of the $\bar \partial$ operator coupled to $V=\CL$ is simply the regularized determinant of the Laplacian 
$\Delta_{\CL}=\bar \partial_{\CL}^{\dagger} \bar \partial_{\CL}$. This has been computed for $g=1$ in \cite{rs} by direct evaluation. For $g>1$ one can regard 
the operator $\bar \partial_{\CL}$ as a family parametrized by the Jacobian $J(\Sigma_g)$, and the analytic torsion can be determined by using the Quillen 
anomaly. An explicit computation has been done in \cite{jj}, but the result gets obscured by the fact that for $g>1$ the 
$\bar \partial_{\CL}$ operator has a nontrivial cokernel of dimension $g-1$. In particular, the result of \cite{jj} 
depends on a choice of a point $p_0 \in \Sigma_g$. The Quillen anomaly gives a closed expression for 
\be
\label{quot}
 {|W(\chi_i(p_0))|^2 \over {\rm det}\, \langle \chi_i| \chi_j\rangle } {\rm det}\, \Delta_{\CL}
\ee
where $\chi_i$, $i=1, \cdots, g-1$ is a basis for $({\rm Coker}\, \bar \partial_\CL)^{*}\simeq {\rm Ker}\, \bar \partial_{\CL^{-1} \otimes K}$, and 
$W$ denotes a Wronskian. The appearance of 
the determinant of the inner products of the $\chi_i$ is typical for computations which use the Quillen anomaly (see for example \cite{agmv}). The result of \cite{jj} for 
(\ref{quot}) is given by
\be
\label{complete}
G({\bf v}, \alpha, \beta)=\Biggl( { {\rm det}' \Delta \over {\rm Vol}(\Sigma_g) {\rm det} \, {\rm Im}\, \tau} \Biggr)^{-{1\over 2}} 
\exp\Bigl[ -2 \pi {\rm Im} {\bf v} ({\rm Im}\, \tau)^{-1}   {\rm Im} {\bf v} \Bigr] \Bigl| \vartheta \Bigl[ {\alpha \atop \beta}\Bigr] ({\bf v}|\tau) \Bigr|^2,
\ee
times a nontrivial function of $p_0$ and $\tau$. In (\ref{complete}), $\Delta$ is the Laplacian on scalars, and as usual the $'$ means that 
we remove the zero modes. $(\alpha, \beta)$ is a characteristic induced by the choice of basepoint $p_0$ through
\be
\CK_{p_0} =\tau \alpha + \beta, 
\ee
where $\CK_{p_0}$ is the vector of Riemann constants associated to the basepoint $p_0$.  The Quillen anomaly measures the failure of holomorphic factorization of the Ray--Singer torsion 
 and reads
 \be
 \label{quillen}
 \bar \partial_{\bf v} \partial_{\bf v}  {  {\rm det}\, \Delta_{\CL} \over {\rm det}\, \langle \chi_i| \chi_j\rangle }= 2\pi \ri g^{i \bar k} dv_i \wedge d{\bar v}_{\bar k}.
 \ee

For $g=1$ the result for the Ray--Singer torsion is much simpler since the dependence on the basepoint 
$p_0$ disappears, and one finds \cite{rs}
\be
\label{gonedet}
 {\rm det}\, \Delta_{\CL}=G({\bf v}, 1/2, 1/2) =\exp\Bigl[ -2 \pi {\rm Im}{\bf v} ({\rm Im}\, \tau)^{-1}  {\rm Im} {\bf v}\Bigr] \Biggl| {\vartheta \Bigl[ {1/ 2  \atop 1/2 }\Bigr] ({\bf v}|\tau) \over \eta(\tau) }\Biggr|^2,
 \ee
 which indeed exhibits an odd characteristic. Therefore, for $g=1$, the quantity
\be
\label{completion}
\widetilde A (p,q)=\log \, {\rm det}\, \Delta_{\CL} -\log |p-q|^2
\ee
provides a modular invariant completion of (\ref{anntheta}) which is antiholomorphic in both the closed moduli $\tau$ and the open moduli 
${\bf v}$. The Quillen anomaly (\ref{quillen}) gives a holomorphic anomaly equation for the {\it open} moduli. 

For $g>1$, the relation between (\ref{anntheta}) and the Ray--Singer torsion is more complicated due to the presence of zero modes. If one 
chooses the basepoint $p_0$ in such a way that the characteristic in (\ref{complete}) is odd, then we can use the Ray--Singer torsion (or rather the quotient (\ref{quot})) as the natural 
completion of $A(p,q)$. Notice that a natural generalization of (\ref{completion}) for $g>1$ is simply to set 
\be
\widetilde A (p,q)=\log \, G({\bf v}, \alpha, \beta)-\log |p-q|^2
\ee
where $G({\bf v}, \alpha, \beta)$ is defined in (\ref{complete}) and we take $(\alpha, \beta)$ to be an odd characteristic. This is a modular invariant 
quantity which satisfies the Quillen anomaly (\ref{quillen}) and has (\ref{anntheta}) as its holomorphic piece. As  
shown in \cite{agmv}, $G({\bf v}, \alpha, \beta)$ can be regarded as the determinant of the Dirac operator on $\Sigma_g$ coupled 
to the flat bundle $\CL$, and for a choice of odd spin structure. This suggests that when considering D-branes on $\Sigma_g$, 
the natural vector bundle $V$ associated to them is not just $\CL$, but $S^{-1} \otimes \CL$, where $S$ is the spin bundle associated to an odd characteristic. This is also consistent with the fact that 
D-branes behave as fermions \cite{adkmv}.

\subsection{Other topological string amplitudes from matrix models}

As we have seen, the annulus amplitude is particularly useful since it can be expressed simply in terms 
of the positions of the branch points of the spectral curve. The main result of \cite{ackm,akemann,eynard,ce, eo} is that 
all free energies and correlation functions in the $1/N$ expansion of one and two--matrix models can be obtained recursively from the 
knowledge of the spectral curve and the two--point function (\ref{twopoint}). The recursion relation of \cite{eynard, eo, eco,eo} involves evaluation of 
residues of meromorphic forms on the curve (\ref{ycurve}) at the branch points. In order to write down this recursion, we first notice that 
near any branchpoint $x_i$ there are two points $q, \bar q$ with the same 
coordinate $x(q)=x(\bar q)$. We now introduce the one--form,
\be
\rd E_q(p)={1\over 2} \int_{ q}^{\bar q} \rd \xi\, B(\xi, p), 
\ee
where $B(\xi,q)$ is the Bergmann kernel (\ref{bergmann}). The recursion relation for the correlation functions reads, 
\be
\ba
&W_g(p, p_1, \cdots, p_h) =\sum_{i=1}^{2s} {\rm Res}_{q=x_i} {\rd E_{q,\bar q}(p) \over (y(q)-y(\bar q)) \rd x (q)} \biggl(  W_{g-1}(q, \bar q, p_1, \cdots, p_h) 
\\ &+\sum_{\ell=0}^g\sum_{m=0}^h \sum_{\sigma \in S(h)} {1\over m! (h-m!)} W_{g-\ell}(q, p_{\sigma(1)}, \cdots, p_{\sigma(m)}) 
W_{\ell}(\bar q, p_{\sigma(m+1)}, \cdots, p_{\sigma(h)})\biggr),
\ea
\ee
where the sum over $\sigma \in S(h)$ is over permutations of $h$ elements. There is also an explicit expression which computes the amplitudes $F_g$ in the 
closed sector from the open amplitudes $W_g(p)$. In that sense, open string amplitudes are more fundamental. 
The resulting formulae for the amplitudes require only the following information about the 
spectral curve: 
\be
x_i , \quad {\rd^k M (x_i) \over \rd p^k}, \quad i=1, \cdots, n, \quad k\ge 0,
\ee
i.e. the position of the branch points and the ``moments" of the function $M(p)$. In principle, the computation 
of these amplitudes is straightforward, but explicit expressions become cumbersome very quickly, specially when the number of cuts $s>1$. We list some of them here,  since 
they will be used in the tests performed in this paper. 

The three--point function at genus zero can be written for general $s$ and was derived in \cite{eynard}. Define first, 
\be
G_0(p,q)= \sigma(p) \sigma (q) \biggl( 4 W_0(p,q) + {2\over (p-q)^2}\biggr). 
\ee
Then, we have
\be
\label{threehole}
W_0(p,q,r)=-{1\over 8 \sigma(p) \sigma (q) \sigma (r)} \sum_{i=1}^{2 s} { G_0(x_i, p) G_0(x_i, q)G_0(x_i, r) \over M(x_i) \sigma'(x_i)^2}.
\ee
For $s=1$ this simplifies to
\be
W_0(p,q,r)={x_1-x_2 \over 8\bigl(\sigma(p) \sigma(q) \sigma(r)\bigr)^{3/2} } \biggl( {(p-x_1)(q-x_1)(r-x_1) \over M(x_1)} -  {(p-x_2)(q-x_2)(r-x_2) \over M(x_2)}\biggr).
\ee
%
%When expanded in $p,q,r$, the first contribution reads
%
%\be
%F_{1,1,1}(t_i)={(x_1-x_2)(M_2 -M_1) \over M_1 M_2}.
%\ee
%
One can also write down results at higher genus. Knowledge of the first moments $M(x_i)$ makes possible to write down an explicit expression for $F_1$. General formulae have been 
written down in \cite{dst,chekhov}. For $s=1$ we have the simple formula, 
\be
\label{gonecut}
F_1=-{1\over 24} \log \bigl[ M(x_1) M(x_2) (x_1 -x_2)^4\bigr], 
\ee
while for $s=2$ we have \cite{akemann,kmt}
\be
\label{genusone}
F_1=-{1\over24}\sum_{i=1}^4\ln M(x_i)-{1\over2}\ln K(k)
-{1\over12}\sum_{i<j}\ln(x_i-x_j)^2
+{1\over8}\ln(x_1-x_3)^2+{1\over8}\ln(x_2-x_4)^2.
\ee
One can also compute explicitly $W_1(p)$, which involves the second set of moments $M'(x_i)$, in the cases $s=1,2$, by using the results of \cite{ackm,akemann}, respectively. 
We emphasize that all these results are particular cases of 
the recursive procedure developed in \cite{eynard, ce, eo} that we just reviewed. 

Although the above expressions were derived in the context of matrix models, we claim that they will lead to the correct open and closed B--model topological string amplitudes 
on a general local Calabi--Yau manifold, provided we use the spectral curve derived from the disk generating functional. In the next two sections we will give supporting evidence for our claim 
by studying some well--known toric CYs. 

\sectiono{Local curves}

\subsection{Closed string sector and phase transitions}

By local curves we refer to the family of toric Calabi--Yau manifolds given by the total 
space of a bundle over a sphere, 
\be
X_p=\CO(p-2) \oplus \CO(-p) \rightarrow \IP^1, \quad p \in \IZ.
\ee
A notational warning: in this section, $p$ will denote the integer labeling $X_p$, and not a point on the spectral curve as in the rest of the sections of this paper. 
Topological string theory on $X_p$ has been very much studied recently (see \cite{aosv} and references therein). As explained in \cite{bp}, the A model 
has to be defined equivariantly. The most natural choice (the equivariant Calabi--Yau case) corresponds to the antidiagonal action on the bundle. As usual, the topological 
closed string amplitudes $F_g(t)$ are put together into a single total free energy, 
\be
\label{totalfgen}
F_{X_p}(g_s,t) =\sum_{g=0}^{\infty} g_s^{2g-2} F^{X_p}_g(t)
\ee
and its exponential is the closed partition function 
\be
Z_{X_p}=\exp \, F_{X_p}(g_s,t).
\ee
Here, $t$ is the only complexified K\"ahler parameter of $X_p$, and it corresponds to the size of $\IP^1$. 

Closed and open partition functions on toric geometries can be computed by using
the topological vertex introduced in \cite{topvertex}, and we will rely extensively on it to test the B--model expressions. We collect here 
some formulae from the theory of the topological vertex which will be useful in the following.  First of all, we define the q--number $[n]$ as
\be
\label{qnumb} [n]=q^{n/2} -q^{-{n/2}}, \qquad q=\re^{g_s}. 
\ee
A representation $R$ of $U(\infty)$ is encoded by a Young tableau, labeled by the lengths of its rows $\{l_i\}$. The
quantity
\be
\ell(R) =\sum_i l_i
\ee
is the total number of boxes of the tableau. Another important quantity associated to a tableau is
\be
\kappa_R=\sum_i l_i(l_i-2i+1).
\ee
We also introduce the quantity
\be
\label{wr}
W_R =q^{-\kappa_R/4} \prod_{\tableau{1} \in R} {1\over [{\rm hook}(\tableau{1})]}, 
\ee
as well as
\be
W_{R_1 R_2}=q^{(\kappa_{R_1} + \kappa_{R_2})/2}
\sum_{R} W_{R_1^t/R} W_{R_2^t/R},
\label{wskew}
\ee
where $R^t$ denotes the transpose tableau (i.e. the tableau where we exchange rows and columns), the ``skew" $W_R$ is defined by
\be
W_{R/R'}=\sum_{R''}N_{R' R''}^R W_{R''}
\ee
and $N_{R_1 R_2}^R$ is the Littlewood--Richardson coefficient defined by the tensor product of $U(\infty)$,
\be
R_1 \otimes R_2 =\sum_R N_{R_1 R_2}^R \, R.
\ee
Both (\ref{wr}) and (\ref{wskew}) are particular cases of the topological vertex, which is given by
\be
\label{vertex}
C_{R_1 R_2 R_3}=
q^{\kappa_{R_2} + \kappa_{R_3}\over 2} \sum_{Q_1,Q_3, Q} N_{Q Q_1}^{~~R_1} N_{Q
Q_3}^{~~R_3^t}
{W_{R_2^t Q_1}W_{R_2 Q_3}\over W_{R_2}}.
\ee

The topological string partition function on $X_p$ is given in terms of the $W_R$ defined in (\ref{wr}):
\be
\label{totalz}
Z_{X_p}=\sum_R W_R W_{R^t} q^{(p-1) \kappa_R/2}Q^{\ell(R)}, \quad Q=(-1)^p \re^{-t}.
\ee
Although (\ref{totalz}) gives an all--genus expression, it is effectively an expansion in powers of $Q$. 
In order to obtain an expression for each $F_g^{X_p}(t)$ to all degrees in $Q$, one usually appeals to mirror symmetry. Since mirror symmetry in this 
equivariant setting is not so well--developed (see however \cite{fj} for recent progress), in \cite{it} we analyzed the saddle of the sum over 
tableaux (\ref{totalz}) directly. We found that the saddle is encoded in a density of tableaux $\rho(\lambda)$ with support on the interval
\be
(x_2,x_1) \cup (x_1, \re), 
\ee
where $x_{1,2}$ are nontrivial functions of the K\"ahler parameter $t$. To specify these, one introduces the mirror map
\be
\label{mirror}
Q=(1-\zeta)^{-p(p-2)} \zeta.
\ee
The endpoints of the cut are given in terms of $\zeta$ by
\be
\label{endpoints}
x_1=(1-\zeta)^{-p}(1+\zeta^{1\over 2})^2, \quad x_2= (1-\zeta)^{-p}(1-\zeta^{1\over 2})^2.
\ee
We note that the variable $\zeta$ is related to the variable $w$ introduced in \cite{it} by $\zeta=1-w$. This saddle--point solution was used in \cite{it} 
to show that the above model has critical behavior for $p>2$. At the critical point 
\be
\label{critical}
\zeta_c={1\over (p-1)^2},
\ee
the theory undergoes a phase transition in the universality class of pure 2d gravity. In fact, if one takes the double--scaling limit 
\be
\zeta \rightarrow \zeta_c, \quad g_s \rightarrow 0, \quad z\, \, {\rm fixed},
\ee
where 
\be
\label{scaled}
z^{5/2} =g_s^{-2} {(p-1)^8 \over 4 (1-\zeta_c)^3}  (\zeta_c-\zeta)^{5},
\ee
then the total free energy (\ref{totalfgen}) becomes the free energy of 2d gravity,
\be
\label{painseries}
F_{(2,3)}(z) =-{4\over 15}z^{5/2}-{1\over 48} \log\, z +\sum_{g\ge 2} a_g z^{-5(g-1)/2},
\ee
where the coefficients $a_g$ can be obtained by solving the Painlev\'e I equation
\be
\label{pone}
u^2 -{1\over 6}u''=z
\ee
satisfied by the specific heat
\be
u(z)=-F''_{(2,3)}(z).
\ee

\subsection{Open string amplitudes}

\begin{figure}[!ht]
\leavevmode
\begin{center}
\epsfysize=4.5cm
\epsfbox{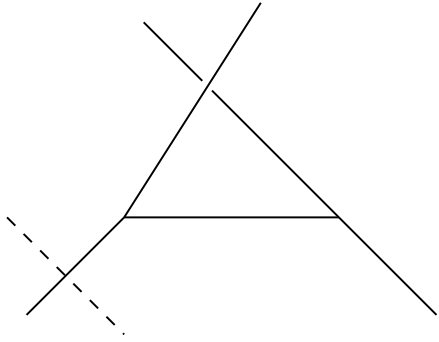}
\end{center}
\caption{The outer brane in $X_p$.}
\label{outercurve}
\end{figure}

We will now consider toric D-branes in the outer leg of the toric diagram associated to the geometry $X_p$, as shown 
in \figref{outercurve}. We will focus here on $p>2$. For $p=1$ (the resolved conifold) there is an explicit matrix model description of the 
background (the Chern--Simons matrix model of \cite{mm}) and one can use this to study D--brane amplitudes \cite{okuyama, hy}. It is well--known that open string amplitudes 
associated to non-compact branes need the specification of a framing \cite{akv}. Here, as well as in the rest of the examples, we will set the framing to zero. 

The open string amplitudes associated to toric D-branes can be computed in the A model by using the topological vertex, and they are encoded in the generating functional 
\be
\label{totalopenf}
F(V)=\log \, Z(V), 
\ee
where 
\be
\label{totalopen}
Z(V)={1\over Z_{X_p}} \sum_R Z_R \, \tr_R\, V.
\ee
In this example we have that
\be
Z_R=\sum_{S} W_{RS} q^{(p-1) \kappa_S/2} Q^{\ell(S)} W_S.
\ee
The functional (\ref{totalopen}) is related to the generating functions (\ref{fgz}) as follows:
\be
F(V)=\sum_{g=0}^{\infty}\sum_{h=1}^{\infty} g_s^{2g-2+h} F_g (z_1, \cdots, z_h), 
\ee
after identifying 
\be
z_1^{w_1} \cdots z_h^{w_h} \leftrightarrow \tr\, V^{w_1} \cdots \tr \, V^{w_h}.
\ee

According to the general philosophy explained in the previous section, the first step in order to compute the open string amplitudes is 
the identification of the spectral curve. This can be obtained from the disk amplitudes. If we had a well--defined mirror symmetry setting for local curves 
we could derive the superpotential with the techniques of \cite{av, akv}. However, here we do not know of any construction like (\ref{curve}) which allows to determine 
$W(z)$ by integration, and we have to use another approach. Fortunately, the analysis of \cite{it} gives a detailed description of the saddle of the sum over partitions in terms 
of an effective matrix model, and we should expect that the resolvent of such a model is the object we need. 
Indeed, let us define 
\be
\omega_0(\lambda)= \int_{x_2}^{\re}\frac{d v}{v}~
\frac{\rho(v)}{\lambda-v}-{1\over \lambda}\log {\lambda\over \lambda -\re}. 
\ee
Notice that this differs from the resolvent as defined in \cite{it}, but it is more natural since it integrates the density of tableaux over its full support. 
After some calculations one finds, 
\be
\label{reslocal}
\omega_0(\lambda)={1\over \lambda} +  {p\over 2 \lambda} \log \biggl[2 { {\sqrt {(\lambda-x_1)(\lambda-x_2)}}-\lambda - {\sqrt {x_1 x_2}} \over  ({\sqrt {x_1 }} +{\sqrt{x_2}})^2}\biggr] ^2+ 
{1\over \lambda} \log \biggl[ {({\sqrt {\lambda-x_1}} +{\sqrt {\lambda-x_2}})^2 \over 4 \lambda}\biggr].
\ee
We can write this, using (\ref{logtanhidentity}), in the form
\be
\omega_0(\lambda) = {1\over 2}(V'(\lambda) + y(\lambda)),
\ee
where 
\be
V'(\lambda)= {p-2\over \lambda} \log \, \lambda -{t\over \lambda}
\ee
plays the role of the potential. The spectral curve is
\be
\label{scapart}
y(\lambda)={2 \over \lambda}\biggl(  \tanh^{-1}\biggl[ {{\sqrt {(\lambda-x_1)(\lambda-x_2)}}\over \lambda- {x_1+x_2\over 2} }\biggr]- p 
\tanh^{-1} \biggl[ {{\sqrt {(\lambda-x_1)(\lambda-x_2)}}\over \lambda+ {\sqrt {x_1 x_2}} }\biggr] \biggr).
\ee
One can check that the resolvent (\ref{reslocal}) is the generating function for disk amplitudes. After using (\ref{resdisk}) one finds, 
\be
\ba
W(z)&=\biggl(1 + Q + \frac{\left( -1 + p \right) \,p\,Q^2}{2} + \frac{p\,\left( -1 + 6\,p - 8\,p^2 + 3\,p^3 \right) \,Q^3}{6} +\cdots \biggr) z \\
  &+ \biggl( \frac{1}{4} + Q + \frac{\left( 1 + 2\,p^2 \right) \,Q^2}{4} + \frac{p\,\left( -2 + 3\,p - 4\,p^2 + 3\,p^3 \right) \,Q^3}{6}  +\cdots\biggr) z^2 + \cdots.
  \ea
  \ee
This is precisely what one obtains in a topological vertex calculation (and leads to integer disk invariants for all $p$ after taking into account the sign $(-1)^p$ 
in (\ref{totalz})). This also confirms that (\ref{scapart}) is the spectral curve for this model. 

As we explained in the previous section, all open and closed string amplitudes in this background can be derived from the knowledge of the branch points and of 
the moments of $M(\lambda)$, which involve the values of $M(\lambda)$ and its derivatives at these branchpoints. The branch points of the curve (\ref{scapart}) 
are of course $x_1, x_2$, and $M(\lambda)$ is obtained 
from the defining equation (\ref{ycurve}). Using this information, together with the mirror map (\ref{mirror}), we 
can compute the annulus amplitude from (\ref{onecutannulus}). We obtain the 
expansion
\be
\ba
A(z_1, z_2)= &\Bigl( Q + p^2 \, Q^2+ {p^2\over 2} \,\left( 1 - 4\,p + 3\,p^2 \right)\, Q^3 \\ 
& + \frac{p^2}{3} \,\left( 1 - 9\,p + 24\,p^2 - 24\,p^3 + 8\,p^4 \right) Q^4 +\cdots\Bigr) z_1 z_2 \\ &
+ \Bigl( Q+(1 + p + p^2)\, Q^2 + \frac{p}{2} \,\left( -1 + 2\,p + 3\,p^3 \right)\, Q^3 \\ 
& \, \, \, + \frac{p}{6}\,\left( -1 + 8\,p - 20\,p^2 + 21\,p^3 - 24\,p^4 + 16\,p^5 \right) \, Q^4 + \cdots\Bigr) z_1 z_2^2 + \cdots
\ea
\ee
in agreement with a vertex computation for the outer brane. It is also easy to compute (\ref{gonecut}). The first moments are given by 
\be
M(x_1)= {2\over x_1} \biggl( {2\over x_1-x_2} -{p \over {\sqrt {x_1}} ({\sqrt {x_1}} + {\sqrt {x_2}})}\biggr), 
\quad M(x_2)= {2\over x_2} \biggl( {2\over x_2-x_1} -{p \over {\sqrt {x_2}} ({\sqrt {x_1}} + {\sqrt {x_2}})}\biggr).
\ee
One finds, up to an irrelevant numerical constant, 
\be
F_1=-{1\over 24} \log \biggl[ { (p-1)^2 \zeta (\zeta_c-\zeta) \over  (1-\zeta)^3} \biggr].
\ee
This genus one amplitude can be written as
\be
F_1=F_1^{\rm inst} +{1\over 24} \log \, Q, 
\ee
where $F_1^{\rm inst}$ is the instanton part of $F_1$ which was presented in eq. (6.1) of \cite{it} and rederived in \cite{fj}. Of course, this 
reproduces the answer obtained from (\ref{totalz}). We have explicitly checked 
that the open string amplitudes with $g=0, h=3$ and $g=h=1$ derived in the matrix model formalism agree with vertex computations. Notice that the 
$g=h=1$ amplitude involves the derivative of $M(\lambda)$ at the branch points, and therefore it probes in detail the structure of the spectral curve (\ref{scapart}).

\subsection{FZZT branes, instantons, and large order behavior}

As we reviewed above, for $p>2$, topological string theory on $X_p$ leads to 2d gravity in a suitable double--scaling limit around the 
transition point in K\"ahler moduli space $\zeta=\zeta_c$. We showed this in \cite{it} by analyzing the closed string amplitudes. Now we have 
closed expressions for open string amplitudes (in particular, for the disk and the annulus), and we can analyze the double--scaling limit 
from the open string sector point of view. 

First of all, we recall that the critical point has a natural interpretation in terms of the spectral curve describing the model: it corresponds precisely to the 
value of the parameter $\zeta$ (or $Q$) for which a zero of the function $M(\lambda)$ coincides with the endpoint 
of the cut $x_1$. This happens when 
\be
M(x_1)=0, 
\ee
which indeed holds at the critical point (\ref{critical}). 

Let us now consider the double--scaling limit of the disk and the annulus amplitude. For the disk 
amplitude, the scaling part is precisely the spectral curve (\ref{scapart}). We first notice that near criticality the endpoints behave as
\be
x_1 =x_1^{(c)}+ \theta (\zeta-\zeta_c)+\CO(\zeta-\zeta_c)^2, \quad x_2=x_2^{(c)}+\CO(\zeta-\zeta_c)^2
\ee
where
\be
\ba
x_1^{(c)}&=(1-\zeta_c)^{-p} {p^2 \over (p-1)^2},\\
x_2^{(c)}&=(1-\zeta_c)^{-p} {(p-2)^2 \over (p-1)^2},\\
\theta&=2 (1-\zeta_c)^{-p} {p(p-1) \over p-2}.\\
\ea
\ee
The open string modulus $\lambda$ in (\ref{scapart}) must scale like
\be
\label{renormopen}
\lambda=x_1^{(c)} -\theta (\zeta-\zeta_c)s+\CO(\zeta-\zeta_c)^2.
\ee
This defines the ``renormalized" open string modulus $s$ and says that, in the scaling limit, the brane must be located near the 
endpoint of the cut. We now reexpress (\ref{scapart}) in terms of the scaling variables $s$ and $z$, which is given in (\ref{scaled}). After some cancellations, we find 
that
\be
\label{fzzt}
{1\over g_s} y(\lambda) \rd \lambda \rightarrow y(s) \rd s=-{4 {\sqrt 2}\over 3} z^{5\over 4}   (2s-1) {\sqrt {1 + s}}\, \rd s,
\ee
for all $p>2$. The r.h.s. is nothing but the Laplace transform of the macroscopic loop operator of 2d gravity \cite{gm,dfgz}, and it corresponds to the disk amplitude of the FZZT brane 
of Liouville theory (see \cite{ss,mmss} and references therein). Notice in particular that, up to a normalization, the double--scaled spectral curve (\ref{fzzt}) 
is nothing but 
\be
T_2(y)=T_3(s),
\ee
where $T_n(\cos z) =\cos (n z)$ are Chebyshev polynomials of the first kind. This is the spectral curve of the $(2,3)$ model \cite{dfgz,ss}. 
The main conclusion of this analysis is that the outer, toric D--brane of the local curve $X_p$ shown in \figref{outercurve} becomes, 
in the double--scaling limit, the FZZT brane of pure 2d gravity. One can further check this by taking the double--scaling limit of the annulus amplitude, which is 
\be
\label{twoloop}
W(\lambda,\mu)\rightarrow {1\over (s-t)^2} \biggl( {1\over 2} {2+ s+t \over {\sqrt {(1+s)(1+t)}}} -1\biggr),
\ee
up to an overall constant. Here we have expressed $\mu$ in terms of the renormarlized open string modulus $t$ (not to be confused with the K\"ahler parameter). The 
r.h.s. of (\ref{twoloop}) is the Laplace transfom of the two--point function of the macroscopic loop operator \cite{ajm,gm}, which 
corresponds to the annulus amplitude of the FZZT brane. Of 
course, the fact that the annulus amplitude (\ref{onecutannulus}) has this limit when we scale its arguments in the way we have done it was noticed long ago in \cite{ajm}. 
What is not obvious here is that annulus amplitudes for toric D--branes are given as well by an expression like (\ref{onecutannulus}). This is a consequence of  
the formalism proposed in section 2, and it implies that the toric brane becomes a FZZT brane near $\zeta=\zeta_c$. 

It can be easily shown from the Painlev\'e I equation (\ref{pone}) that the coefficients $a_g$ of the 
perturbative genus expansion (\ref{painseries}) have the asymptotics
\be
\label{aglarge}
a_g \sim A^{-2g} (2g)!, 
\ee
where 
\be
\label{instantonaction}
A={8 {\sqrt 3 }\over 5},
\ee
see for example \cite{dfgz}\footnote{The normalization in section 7.2 of this reference is appropriate for an even potential. The generic critical potential 
leads to an extra ${\sqrt {2}}$ in $A$ as compared to their computation.}. Based on general considerations in quantum field theory \cite{zj} we 
would expect that the large order behavior in (\ref{aglarge}) is controlled by instanton effects of the form 
\be
\exp\bigl( -A z^{5\over 4}\bigr).
\ee
It was noticed in \cite{shenker, david} that this instanton has a natural interpretation in terms of eigenvalue tunneling. Recall that in the context of the matrix models, 
the variable $x$ in (\ref{ycurve}) is an eigenvalue of the underlying matrix. The instanton effect is obtained as we move one single eigenvalue at the end of the 
cut $x_1$ to a saddle point of the effective potential, located at the top of a potential barrier. This point $x_0$ satisfies
\be
M(x_0)=0,
\ee
and it is the zero of $M(\lambda)$ which collides with $x_1$ at the critical point. The instanton action is given by  
\be
\label{instoff}
A=\int_{x_1}^{x_0} y(\lambda) \rd \lambda. 
\ee
As we approach the critical point, $x_1 \rightarrow x_0$ and (\ref{instoff}) vanishes. But in the double--scaling limit, since $g_s\rightarrow 0$ simultaneously, we find a nonzero 
result for $A/g_s$. David explicitly showed in \cite{david} that one obtains in this way the value (\ref{instantonaction}). Indeed, we can recover his result from our toric 
realization of 2d gravity. In terms of the renormalized open string variable $s$, the positions of $x_1, x_0$ are given by $s=-1$ and $s=1/2$, respectively, as one 
can immediately verify by looking 
at (\ref{renormopen}) and (\ref{fzzt}). We can then compute the instanton action in the double--scaling limit to be
\be
-{4 {\sqrt 2}\over 3}  \int_{-1}^{1\over2} \rd s\,  (2s-1) {\sqrt {1 + s}}
\ee
which agrees precisely with (\ref{instantonaction}). 

We would like however to extend this analysis {\it off criticality}, in order to understand non-perturbative effects and the large order 
behavior of perturbation theory in topological string theory. In analogy with the 2d gravity limit, the relevant instanton 
effects will be obtained as we move the toric brane form the endpoint of the cut to a saddle point. The instanton action will now be a function of the couplings of the 
theory, as it is familiar from the analysis of instanton effects and large order behavior in the $1/N$ expansion (see \cite{hb} for examples in low dimensions). In 
our case, the action will depend on $Q$, and we will have a non-perturbative effect of the form 
\be
\exp \Bigl( -{A(Q)\over g_s} \Bigr), 
\ee
leading to the large order behavior 
\be
F_g(Q) \sim (A(Q))^{-2g} (2g)!. 
\ee
The instanton action $A(Q)$ is given by the integral (\ref{instoff}) in the full theory, where 
$y(\lambda)$ is (\ref{scapart}) and $x_0$ is now the zero of $M(\lambda)$ which gives the dominant instanton contribution, 
i.e. the instanton with the smallest action in absolute value (recall that in 
general the instanton action might be complex). At least for $Q$ near $Q_c$, we expect that, due to the connection to 2d gravity, 
the point $x_0$ becomes the point $s=1/2$ 
in the double--scaling limit. This means that it will behave like
\be
\label{dsxo}
x_0=x_1^{(c)}-{1\over 2} \theta (\zeta-\zeta_c) +\CO(\zeta-\zeta_c)^2
\ee
near the critical point. The solutions of $M(\lambda)=0$ are easily obtained for any given $p$. For example, for $p=3$, one finds 
a unique solution,
\be
x_0={4 x_1 x_2  \over ({\sqrt {x_1}} - {\sqrt {x_2}})^2}, \quad p=3,
\ee
which behaves indeed like (\ref{dsxo}). For $p=4$ there are three solutions, and the one with the right 
critical behavior (\ref{dsxo}) is given by
\be
x_0={2 {\sqrt {x_1}} x_2\over {\sqrt {x_1}} - {\sqrt {x_2}}}, \quad p=4.
\ee
The integral (\ref{instoff}) can be computed in closed form. One obtains, 
\be
\label{instp}
A(Q)= F(x_0) -F(x_1),
\ee
where
\be
\label{bigf}
\ba
F(x)&= -\log \,(f_1(x)) \biggl( \log \, (f_1(x)) - 2\log \Bigl( 1 + {2 f_1(x) \over ({\sqrt {x_1}}- {\sqrt{x_2}})^2} \Bigr)  
+ \log \Bigl( 1 + {2 f_1(x) \over ({\sqrt {x_1}}+{\sqrt{x_2}})^2} \Bigr)   \biggr)\\
&-2 {\rm Li}_2 \Bigl( -{2 f_1(x) \over ({\sqrt {x_1}}- {\sqrt{x_2}})^2} \Bigr)  -
2 {\rm Li}_2  \Bigl( - {2 f_1(x) \over ({\sqrt {x_1}}+{\sqrt{x_2}})^2} \Bigr) -  \log { (x_1-x_2)^2 \over 4}\log \, x \\
&-p\log \,(f_2(x)) \biggl( \log \, (f_2(x)) +2\log \Bigl( 1 - { f_2(x) \over 2 {\sqrt {x_1 x_2}}} \Bigr)  
- \log \Bigl( 1 - {2 f_2(x) \over ({\sqrt {x_1}}+{\sqrt{x_2}})^2} \Bigr)   \biggr)\\
&-2 p {\rm Li}_2 \Bigl( -{ f_2(x) \over 2 {\sqrt {x_1 x_2}}} \Bigr)  + 2 p
{\rm Li}_2  \Bigl( {2 f_2(x) \over ({\sqrt {x_1}}+{\sqrt{x_2}})^2} \Bigr) +{p\over 2} (\log x)^2 + p \log  ({\sqrt {x_1}}+ {\sqrt{x_2}})^2 \log\, x
\ea
\ee
and 
\be
\ba
f_1(x)&= {\sqrt {(x-x_1)(x-x_2)}} + x-{x_1+x_2\over 2}, \\
f_2(x)&= {\sqrt {(x-x_1)(x-x_2)}} + x+ {\sqrt{x_1 x_2}}.
\ea
\ee
It is easy to check, for the values of $x_0$ that scale like (\ref{dsxo}), that this action indeed becomes (\ref{instantonaction}) in the double-scaling limit. 
\begin{figure}[!ht]
\leavevmode
\begin{center}
\epsfysize=5cm
\epsfbox{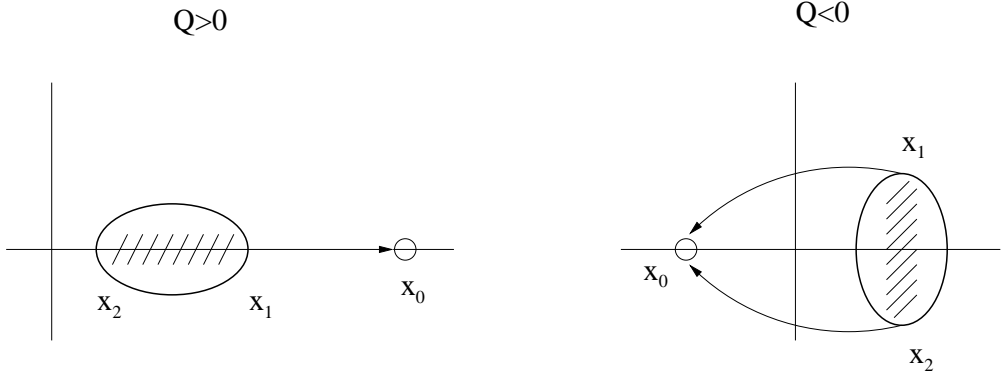}
\end{center}
\caption{This figure shows the instanton effects in the $p=3$ theory for real $Q$. When $Q>0$ the cut $(x_1, x_2)$ is in the real axis 
and the instanton is obtained by moving one brane from $x_1$ to $x_0$. For $Q<0$, $x_{1,2}$ are complex, with $x_1=x_2^*$. 
The cut is parallel to the imaginary axis and there are 
two complex conjugate instantons obtained by moving the branes at the two edges of the cut.}
\label{tunnel}
\end{figure}

\begin{figure}[!ht]
\leavevmode
\begin{center}
\epsfysize=4.5cm
\epsfbox{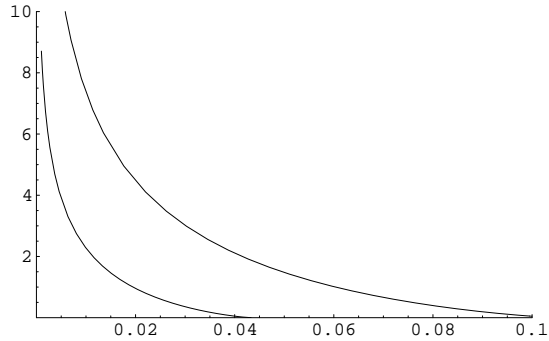}
\end{center}
\caption{This figure shows the instanton action (\ref{instp}) as a function of $Q_c>Q>0$. The upper line corresponds to $p=3$, while the lower line 
is $p=4$. The action vanishes at $Q_c(p)$, and it is real and positive.}
\label{insts}
\end{figure}

Let us first analzye the instanton effects for $Q$ real and positive in the range $Q_c > Q >0$. In this case, the branch points $x_1, x_2$, as well as $x_0$, are on the real axis, 
with $0<x_2<x_1<x_0$, and the instanton effect corresponds to a single brane moving from $x_1$ to $x_0$ (see the left hand side of \figref{tunnel}). 
This leads to an instanton action which is real and positive and vanishes at the critical point $Q_c$. In \figref{insts} we show the values of this action for $p=3, 4$ as a function of $Q$ in this range. 
Since this instanton effect is of order $1/g_s$, this analysis shows that, for $p>2$ and $Q$ real and positive, the genus expansion (\ref{totalfgen}) has the typical stringy divergence $\sim (2g)!$ \cite{shenker} and it is not Borel summable. 

It is interesting to compare this behavior to the one found in the Hermitian matrix model with a potential of the form
\be
{1\over 2} \tr \, M^2 +  \kappa \tr\, M^4.
\ee
The $F_g(\kappa)$ are analytic at $\kappa=0$ with a radius of convergence $r=1/48$. The genus expansion diverges as $(2g)!$, and the instanton action 
behaves very differently depending on the value of $\kappa$. For $-1/48 <\kappa<0$ 
it is real and positive, and the genus expansion is not Borel summable. This of course reflects the instability of the model for negative $\kappa$ and 
leads to the non--Borel summability of 2d gravity. Topological string theory 
on $X_p$ with $p>2$, $Q_c> Q>0$ shows exactly the same behavior that the quartic matrix model in the unstable region.

Of course, in topological string theory the parameter $Q$ is in general complex, due to the presence of the $B$ field, and this will lead to complex instanton actions. 
Let us consider for example the case $p=3$, and $Q$ in the range $-Q_c < Q<0$, in which the genus expansion (\ref{totalfgen}) is still real and each $F^{X_3}_g(Q)$ converges. 
For these values of $Q$ the branch points $x_1, x_2$ are complex conjugate to each other, and the point $x_0$ moves to the negative real axis. The instanton effects are obtained 
by analytic continuation from the case $Q>0$ that we have just analyzed, where the instanton is real. The instanton obtained by moving one brane from $x_1$ to $x_0$ is  
now complex, but since the free energy is still real, we should expect two 
instantons with complex conjugate actions. The other instanton is obtained by moving one brane from the other end of the Fermi surface, i.e. 
from $x_2$, to $x_0$ (see the right hand side of \figref{tunnel}). 
Since the instantons are now complex, the resulting genus expansion might be Borel summable. For general, complex $Q$, the cut $(x_1, x_2)$ will be 
located in a generic direction in the complex plane, and one should analyze in detail the possible instantons obtained by moving the branes to the zeroes of $M(\lambda)$. 
The instanton actions in this more general setting can be computed as well from the analytic result (\ref{bigf}). 

Our analysis indicates that, for real positive values of $Q<Q_c$, the string perturbation series will bot be Borel summable, as in 
2d gravity, but it also suggests that one can find Borel--summable expansions by rotating $Q$ in the complex plane. 
As a final comment, we recall that instanton effects in 2d gravity can be interpreted in terms of ZZ branes \cite{akk}. The contribution of the ZZ brane can be written as the difference of 
two FZZT branes located at the points $x_1$ and $x_0$, as in (\ref{instp}) \cite{ss,mmss}. The above computation in topological string theory is very similar due to 
the underlying matrix model structure. 

\sectiono{Local surfaces}

In this section we discuss two of the most studied examples in topological string theory, namely the local Calabi--Yau manifolds obtained 
by considering the anticanonical bundle on $\IP^1 \times \IP^1$ and on $\IP^2$. We will make explicit the prescription of section 2 to compute open string amplitudes 
and present some examples. We will focus in all cases in outer branes with zero framing. 

\subsection{Local $\IP^1 \times \IP^1$}

The local $\IP^1 \times \IP^1$ geometry has two K\"ahler parameters $Q_s$, $Q_t$. The mirror can be described by the equation (\ref{curve}) with \cite{akv}
\be
F(\re^u, \re^v)= \re^u + z_t \re^{-u} + \re^v + z_s \re^{-v} +1.
\ee
The variables $u$, $v$ are only defined up to a ${\rm SL}(2,\IZ)$ transformation, and one chooses one description or another 
depending on the brane that one wants to describe. The variables $z_t, z_s$ are the linear sigma model parameters, which are 
related to $Q_t$, $Q_s$ through the mirror map. The explicit expression for this map is, 
\be
\label{mirrormapone}
\ba
Q_t&= z_t \exp \biggl[ 2\sum_{k,l\ge 0} {(2k + 2l -1)! \over (k!)^2 (l!)^2 } z_t^k z_s^l \biggr], \\
Q_s&= z_s \exp \biggl[ 2\sum_{k,l\ge 0} {(2k + 2l -1)! \over (k!)^2 (l!)^2 } z_t^k z_s^l \biggr].
\ea
\ee
The inversion of (\ref{mirrormapone}) gives
\be
z_t=Q_t - 2 Q_s Q_t - 2 Q_t^2 + \cdots, \quad z_s=Q_s - 2 Q_s Q_t - 2 Q_s^2 + \cdots.
\ee

\begin{figure}[!ht]
\leavevmode
\begin{center}
\epsfysize=4.5cm
\epsfbox{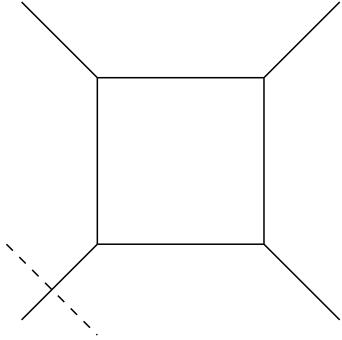}
\end{center}
\caption{The outer brane in local $\IP^1 \times \IP^1$.}
\label{outerpone}
\end{figure}

Let us now consider the disk amplitude for an outer brane in this geometry, shown in \figref{outerpone}. This was computed in \cite{akv,lm}\footnote{This is brane III in the notation of \cite{akv}. The results 
presented in table 3 of this paper do not seem to be correct, however. A computation of the disk superpotential for the outer brane of local $\IP^1 \times \IP^1$ is presented in example 
A.2 of \cite{lm}, and this can be checked against the topological vertex.}. The function $v(z)$ is given by 
\be
v(z)=\log \biggl[ {1- z + z_s z^2 + {\sqrt { (1 - z + z_s z^2)^2 - 4 z_t z^2}} \over 2}\biggr].
\ee
In order to make contact with enumerative results in the A model, we have to take into account the important fact that 
the open string modulus $z$ is not a flat coordinate. There is a nontrivial open mirror map, and in the case of the outer brane in local $\IP^1 \times \IP^1$ this is 
given by 
\be
\label{openmm}
z=  \Bigl( {Q_s \over z_s}\Bigr)^{1\over 2} Z, 
\ee
where $Z$ is the flat coordinate. The quantities that we compute in the B model will be functions of $z_t, z_s,z$, and we will have to reexpress them in terms 
of the flat coordinates $Q_t, Q_s,Z$ in order to compare them with the quantities computed in the A model.

As we explained in section 2, the spectral curve can be read from $v(z)$. In terms of $p=z^{-1}$ we find 
\be
y(p)={2\over p} \tanh^{-1} \biggl[ {{\sqrt{  (z_s - p + p^2)^2 - 4 z_t p^2 }} \over z_s - p + p^2}\biggr].
\ee
This spectral curve has four branch points, given by
\be
x_{1,2}= {1\over 2} - {\sqrt {z_t}} \mp {1\over 2} {\sqrt {(1+ 2  {\sqrt {z_t}})^2 - 4z_s}}, \quad x_{3,4}= {1\over 2} + {\sqrt {z_t}}\pm {1\over 2} {\sqrt {(1-2  {\sqrt {z_t}})^2 - 4z_s}}.
\ee
The labeling of the cuts has been chosen in such a way that the cuts $(x_1, x_4)$, $(x_2,x_3)$ shrink to zero size when $z_t=z_s=0$. 
We can now use the explicit expression (\ref{twocutannulus}) to extract the generating function $A(p,q)$ of annulus amplitudes for outer branes, and we have to reexpress the result in terms 
of the flat coordinates. The open mirror map is the same for both the two open moduli, therefore we will set $p=z_1^{-1}$, $q=z_2^{-1}$ and use (\ref{openmm}) for 
both. One obtains the following expansion, 
\be
\ba
&A(Z_1, Z_2) =\Bigl( Q_t + 2 Q_t Q_s + 4 Q_t^2 Q_s + 6 Q_t^3 Q_s + 36 Q_t^2 Q_s^2 + 4 Q_t Q_s^2 + 6 Q_t Q_s^3 + \cdots\Bigr) Z_1 Z_2 \\ 
&+\Bigl( Q_t + Q_t^2 +2 Q_t Q_s + 3 Q_t Q_s^2 + 5 Q_t Q_s^3 + 6 Q_t^2 Q_s + 30 Q_t^2 Q_s^2 + 10 Q_t^3 Q_s +\cdots \Bigr)Z_1 Z_2^2  \\
&+\Bigl( Q_t + 2Q_sQ_t + 3Q_s^2Q_t + 4Q_s^3Q_t + 3Q_t^2 + 12Q_sQ_t^2 + 40Q_s^2Q_t^2 + Q_t^3 + 24Q_sQ_t^3+\cdots\Bigr)  Z_1 Z_2^3 \\
&+\Bigl( Q_t + 2Q_sQ_t + 3Q_s^2Q_t + 4Q_s^3Q_t + \frac{5Q_t^2}{2} + 10Q_sQ_t^2 + 33Q_s^2Q_t^2 + Q_t^3 + 20Q_sQ_t^3 +\cdots \Bigr) Z_1^2Z_2^2  \\
& \,\,\,\,\, +\cdots
\ea
\label{annpone}
\ee
Notice that this satisfies the integrality conditions (\ref{integers}). The above result can be compared to that obtained with the topological vertex. 
The total partition function for this geometry can be written as 
\be
Z_{\IP^1 \times \IP^1}=
\sum_{R_i}Q_t^{\ell(R_1)+ \ell(R_3)} Q_s^{\ell(R_2)+ \ell(R_4)}
q^{\sum_i \kappa_{R_i}/2}  C_{0 R_4 R_1^t} C_{0 R_1 R_2^t} C_{0 R_2 R_3^t}
C_{0 R_3 R_4^t}.
\label{ponebis}
\ee
The generating functional for the open string amplitudes in the presence of an outer brane is given again by (\ref{totalopen}), where now $Z_R$ is given by
\be
Z_R=\sum_{R_i}Q_t^{\ell(R_1)+ \ell(R_3)} Q_s^{\ell(R_2)+ \ell(R_4)}
q^{\sum_i \kappa_{R_i}/2}  C_{R R_4 R_1^t} C_{0 R_1 R_2^t} C_{0 R_2 R_3^t}
C_{0 R_3 R_4^t}.
\ee
One can easily check that the annulus amplitudes computed from these vertex formulae agree with the expansion (\ref{annpone}). 

It is possible to check as well the open, three--hole amplitude at genus zero and the closed, genus one amplitude $F_1$, by using the formulae presented in 
section 2. We need the first moments of the 
function $M(p)$, which are given by
\be
M(x_i) ={2\over x_i (z_s -x_i +x_i^2)}, \quad i=1, \cdots, 4.
\ee
One finds, for example, 
\be
\ba
F_0(Z_1, Z_2, Z_3)=\Bigl(&Q_t + 2Q_sQ_t + 3Q_s^2Q_t + 5Q_s^3Q_t + 7Q_s^4Q_t + Q_t^2 + 6Q_sQ_t^2+ 40Q_s^2Q_t^2 \\
& + 189Q_s^3Q_t^2 + 10Q_sQ_t^3 + 217Q_s^2Q_t^3 + 14Q_sQ_t^4
+\cdots\Bigr) Z_1 Z_2 Z_3 +\cdots,
\ea
\ee
while for the genus one, closed string amplitude we find from (\ref{genusone}) the expansion 
\be
\ba
F_1(Q_s, Q_t)=&-{1\over 12} \log Q_t -\frac{Q_s}{6} - \frac{Q_s^2}{12} - \frac{Q_s^3}{18} - \frac{Q_s^4}{24} - \frac{Q_t}{6} - \frac{Q_s\,Q_t}{3} - \frac{Q_s^2\,Q_t}{2} - \frac{2\,Q_s^3\,Q_t}{3} - \frac{Q_t^2}{12}
\\ & - 
  \frac{Q_s\,Q_t^2}{2}  + \frac{37\,Q_s^2\,Q_t^2}{6} 
 - \frac{Q_t^3}{18} - \frac{2\,Q_s\,Q_t^3}{3} - \frac{Q_t^4}{24} +\cdots.
  \ea
  \ee
  The worldsheet instanton contributions in this quantity are indeed the correct ones, but the classical part $-\log \, Q_t/12$ is asymmetric in that it only includes the $t$ parameter. This is probably due to the fact that the disk amplitude we start with is itself asymmetric in its treatment of $Q_t, Q_s$. In fact, it is nice that this asymmetry does not 
  affect the instanton part of $F_1$, which is correctly reproduced. 

This ends our tests for local $\IP^1 \times \IP^1$. We will now briefly consider the case 
of local $\IP^2$.

\subsection{Local $\IP^2$}

The local $\IP^2$ geometry has one K\"ahler parameter $Q$. The mirror can be described by 
\be
F(\re^u, \re^v)= -\re^v + \re^u +1 -z_t \re^{3u-v},
\ee
where $z_t$ is the linear sigma model parameter.
The mirror map is given by
\be
\label{mirrormaptwo}
Q= z_t \exp \biggl[ 3\sum_{k\ge 0} {(3k-1)! \over (k!)^3 } (-1)^k z_t^k  \biggr],
\ee
and its inversion reads
\be
z_t=Q+ 6\,Q ^2 + 9\,Q ^3 + 56\,Q ^4 - 300\,Q ^5 + 3942\,Q^6+ \cdots.
\ee

\begin{figure}[!ht]
\leavevmode
\begin{center}
\epsfysize=4.5cm
\epsfbox{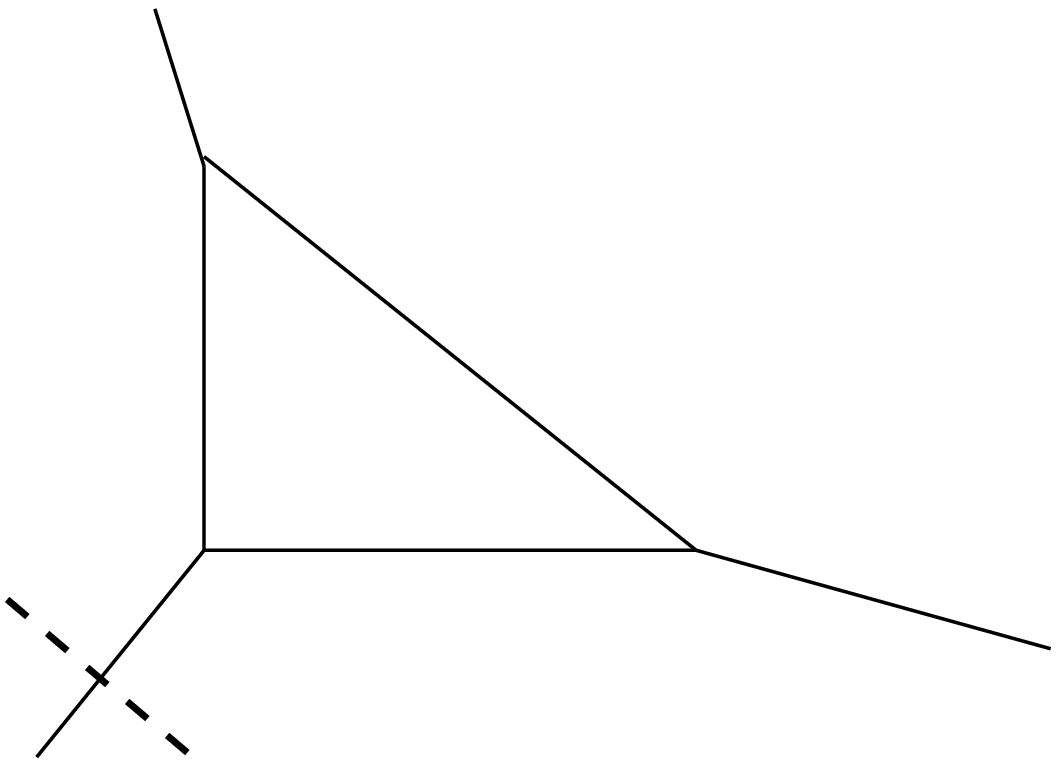}
\end{center}
\caption{The outer brane in local $\IP^2$.}
\label{outerptwo}
\end{figure}

Let us now consider the disk amplitude for an outer brane in this geometry, shown in \figref{outerptwo}. This has been computed in \cite{akv,lm}. The function $v(z)$ is given by 
\be
v(z)=\log \biggl[ {1+z + {\sqrt { (1 +z)^2 -4 z_t z^3}} \over 2}\biggr].
\ee
The open mirror map for the outer brane is given by
\be
z=  \Bigl( {Q \over z_t}\Bigr)^{1\over 3} Z, 
\ee
where $Z$ is the flat coordinate. After setting $z=p^{-1}$, we find the spectral curve
\be
\label{ptwosc}
y={2\over p} \tanh^{-1} \Biggl[ {{\sqrt {p\bigl(p(p +1)^2 - 4 z_t \bigr)}} \over p(p+1)}\Biggr].
\ee
The branch points of the curve are $x_1=0$ plus the roots of the cubic equation
\be
x(x+1)^2-4 z_t=0.
\ee
In terms of
\be
\xi=\Bigl( 1 + 54\, z_t + 6\,{\sqrt{3}}\,{\sqrt{z_t\,\left( 1 + 27\,z_t\right) }}\Bigr)^{1\over 3}
\ee
they are given by
\be
x_2=-{2\over 3} + {1 \over 3} \Bigl(\omega  \xi+{1\over \omega \xi}\Bigr), \quad x_3=-{2\over 3} + {1 \over 3} \Bigl(\omega^* \xi +{1\over \omega^* \xi}\Bigr), \quad 
x_4={(\xi-1)^2 \over 3\xi},
\ee
where $\omega=\exp(2 \ri \pi/3)$. 

We can now compute the annulus amplitude for the outer brane. We find the expansion, 
\be
\ba
A(Z_1, Z_2) &=\Bigl(-Q + 4\,Q^2 - 35\,Q^3 + 400\,Q^4 - 5187\,Q^5  + \cdots\Bigr) Z_1 Z_2 \\ 
&+\Bigl( Q - 3\,Q^2 + 24\,Q^3 - 261\,Q^4 + 3288\,Q^5 +\cdots \Bigr)Z_1 Z_2^2  \\
&+\Bigl( -Q + 4\,Q^2 - 25\,Q^3 + 248\,Q^4 - 2970\,Q^5+\cdots\Bigr)  Z_1 Z_2^3 \\
&+\Bigl( -Q + \frac{7\,Q^2}{2} - 20\,Q^3 + 194\,Q^4 - 2299\,Q^5+\cdots \Bigr) Z_1^2Z_2^2  +\cdots,
\ea
\label{annptwo}
\ee
which again is in agreement with the results for the outer brane amplitudes obtained from the topological vertex. In this case, the total 
partition function is
\be
Z_{\IP^2}=\sum_{R_1,R_2,R_3}
(-1)^{\sum_i \ell(R_i)}Q^{\sum_i \ell(R_i)}
q^{-\sum_i \kappa_{R_i}}C_{0 R_2^t R_3} C_{0 R_1^t R_2}
C_{0 R_3^t R_1},
\label{ptwo}
\ee
and the amplitude for the outer brane with zero framing is obtained from 
\be
Z_{R}=\sum_{R_1,R_2,R_3}
(-1)^{\sum_i \ell(R_i)}Q^{\sum_i \ell(R_i)}
q^{-\sum_i \kappa_{R_i}}C_{R R_2^t R_3} C_{0 R_1^t R_2}
C_{0 R_3^t R_1},
\ee
We can again compute the three--hole, genus zero amplitude. The first moments are given by
\be
M(x_i)={2 \over x_i^2(x_i+1)}, \quad i=1, \cdots, 4.
\ee
Notice that $M(p)$ diverges at the first branch point $x_1=0$. This is no problem for the three--hole amplitude (\ref{threehole}), and simply means that the first term 
in the sum vanishes. One finds, 
\be
F_3(Z_1, Z_2, Z_3)= \Bigl( -Q + 3\,Q^2 - 36\,Q^3 + 531\,Q^4 - 8472\,Q^5 +\cdots\Bigr) Z_1 Z_2 Z_3 +\cdots,
\ee
again in agreement with a vertex computation. Of course, for the expression (\ref{genusone}) we should subtract the diverging moment for $x_1=0$. Once this is done, 
one obtains the right expression for $F_1$. This ends the tests of the formalism proposed in this paper. 

A final comment on this example is in order. It is well--known that local $\IP^2$ has a conifold point where closed topological string amplitudes diverge. This point 
corresponds to the critical value 
\be
z_t^{(c)}=-{1\over 27}.
\ee
In terms of the spectral curve (\ref{ptwosc}), this is described 
by the collision of $x_3$ and $x_4$, and it leads to a critical behavior that has been studied for example in \cite{akemann, birthcut}. On 
the other hand, based on general properties of topological string theory, we should expect that 
the critical theory at this point is the $c=1$ string at self--dual radius \cite{gv}. It was indeed suggested in 
\cite{moore,dgkv} that the universality class associated to two colliding cuts 
should be indeed the $c=1$ string, and some aspects of this correspondence have been further clarified in the 
series of papers \cite{bertoldi}. It would be very interesting to analyze the open sector of topological strings on local $\IP^2$ near criticality, as we did in section 3 for local curves. 
This would make possible to extract nonperturbative effects due to D-branes and in particular to determine the large order behavior of the perturbative string expansion.  
 
\sectiono{Conclusions and open problems}

In this paper we have proposed a formalism inspired on matrix models which allows us to compute in closed form open and closed string amplitudes 
of the type B topological string on toric backgrounds. This formalism exploits the result of \cite{topvertex,adkmv} which identifies the local B--model with the theory of a 
quantum chiral boson living on a Riemann surface. Since the general solution of matrix models on a spectral curve can be identified with such 
a quantum theory, we conclude that the rich results developed in the study of loop equations of matrix models can be applied in this context. 

Notice that the ``effective" matrix model description introduced here is rather different from the multi--matrix models of \cite{m,akmv}. Those models are 
parametrized directly by the flat coordinates, which appear either in the action or through the ranks of the matrix fields. 
In the formalism of this paper, the parameters entering into the description are not flat coordinates, and one has to use the 
mirror map (open and closed) to translate the answers into a suitable form for 
comparing with the A model. 

Of course, the results of Dijkgraaf and Vafa in \cite{dv} provided a direct conection between matrix models and Calabi--Yau geometry, but the backgrounds 
analyzed in \cite{dv} are in a sense orthogonal to the ones that appear in Gromov--Witten theory and in the theory of the topological vertex. Here we have seen how 
to fit the latter in the matrix model formalism, and we have tested these ideas in various examples. The case of local curves is particularly interesting due to 
the connection of these backgrounds with 2d gravity. This makes possible to find the instanton effects responsible for the large order behavior of topological string theory 
on those manifolds. 

As a corollary of this matrix description we have a picture in which the toric Calabi--Yau geometry emerges as a large $N$ description 
of a fundamentally discrete system of $N$ interacting eigenvalues, and with a characteristic length of order $1/N \sim g_s$. This picture has already 
appeared in many systems which exhibit a gauge/gravity duality, in particular in the Dijkgraaf--Vafa scenario \cite{dv}. The results of this paper extend it to 
toric Calabi--Yau backgrounds. The discretization of the geometry that emerges is similar to that suggested in \cite{orv}, where a crystal model with characteristic length 
$g_s$ was proposed as the underlying stringy description of some toric geometries. It would be interesting to understand better the connection between these two pictures. 

Let us finish this paper with a list of open problems and directions for further research. 

\begin{itemize}

\item From a mathematical point of view, the results of this paper offer a wealth of mirror conjectures about the open string sector of 
toric Calabi--Yau manifolds. It should be possible to generalize the analysis of \cite{gz}, which proved a mirror conjecture for the 
disk ampitude, to the case of annulus amplitudes and beyond. Of course, if our formalism holds in general, one should be able to prove direclty in Gromov--Witten theory 
that the generating functionals of open (or relative) Gromov--Witten invariants satisfy the recursion relations typical of loop equations found in \cite{eynard,eo}. Since these invariants 
are given by sums over partitions, our formalism is equivalent to asserting that the corrections to the saddle--point 
of these sums satisfy the loop equations of an ``effective" matrix model, at least for this class of examples. 

\item More tests of our formalism are needed in order to understand 
amplitudes with higher $g$ and $h$, and in particular the implications for the closed string sector. But more 
importantly, in its current form this formalism only applies to amplitudes for outer branes, and one should generalize it 
to general configurations of branes in these geometries. A hint to understand this is given by the following observation: one can analyze the 
disk amplitude in terms of an open generalization of Picard--Fuchs equations \cite{lm,lmw,india}, and it is known that outer and inner branes are on different 
sides of a phase transition which moreover involves different mirror maps. We expect that the same universal expressions in terms of non--flat 
coordinates (like (\ref{twocutannulus})) hold for all branes, but that one has to use different open mirror maps in different phases, 
leading in this way to the different expressions for open string amplitudes that one finds in the A model. After all, in the B--model 
outer and inner branes should be smoothly connected, since the Riemann surface described by (\ref{curve}) smooths out the toric diagram of the A model\footnote{This issue has 
been emphasized and discussed in \cite{amir}.}. The mirror map for inner branes mixes open and closed coordinates in a nontrivial way \cite{lm}, and one should 
probably work out a more general Picard--Fuchs system involving both the closed moduli and the $h$ open moduli $z_i$ appearing in the general amplitude (\ref{fgz}). 

\item As we mentioned in the introduction, it is not known what is the analogue of the holomorphic anomaly equations for open strings. However, by using the 
results of \cite{eo}, it is possible \cite{em} to find the holomorphic anomaly equations satisfied by the functions $W_g(z_1, \cdots , z_h)$ defined by the recursive 
procedure of \cite{eynard, eco,eo} (the holomorphic anomaly equation for the annulus amplitude is written down in (\ref{holan})). It is likely that these equations, which are in principle
valid in the local case, give a strong hint of the structure of the holomorphic anomaly equations for open strings in the general case. Of course, one advantage of the 
procedure of this paper is that it is a recursive B--model calculation which does not exhibit the ambiguity of \cite{bcov}. The holomorphic 
anomaly equations can be used to determine closed string amplitudes on local Calabi--Yau manifolds \cite{kz,hk}, and it has been recently shown that 
for these backgrounds the holomorphic ambiguity can be fixed from a ``gap" behavior at the conifold point \cite{hk}. It would be interesting to compare the formalism 
of this paper in the closed case with the procedure of \cite{kz,hk}, and to explore in more detail the modular properties of the open string amplitudes.

\item From the results of \cite{it} and of section 3 of this paper, it seems that a way to understand certain nonperturbative aspects of topological string theory is to 
relate them to noncritical string theories, where these aspects are better understood. When discussing local $\IP^2$ we mentioned that there should be a double--scaling limit 
of the spectral curve (\ref{ptwosc}) which gives the $c=1$ string. A better understanding of this limit would shed new light 
on nonperturbative aspects of topological string theory on local $\IP^2$, like for example the identification of instanton configurations. In the case of $\IP^1 \times \IP^1$, there is 
besides the $c=1$ limit a geometric engineering regime which leads to the $SU(2)$ Seiberg--Witten theory \cite{kkv}. The 
specialization of our formalism to this regime would give an effective matrix model description of Nekrasov's instanton expansion and its gravitational corrections \cite{nekrasov}. 
That such a matrix model description should exist has been also pointed out in \cite{bertoldi}.

\end{itemize}

\section*{Acknowledgments}

I would like to thank Gernot Akemann, Luis \'Alvarez--Gaum\'e, Nicola Caporaso, Bertrand Eynard, C\'esar G\'omez, Luca Griguolo, Amir Kashani--Poor, Volodya Kazakov, 
Albrecht Klemm, Ivan Kostov, Wolfgang Lerche, Sara Pasquetti, 
Domenico Seminara and Eric Zaslow for useful discussions.

\end{document}